\documentclass[pra,twocolumn,showpacs,superscriptaddress]{revtex4}
\usepackage{epsfig,graphicx,amssymb,amsmath,color,bm}
\usepackage{subfigure}
\usepackage{setspace}

\newcommand{\eref}[1]{Eq.~(\ref{#1})}

\begin{document}

\title{Optomechanical cooling in the non-Markovian regime}
\author{Wen-Zhao Zhang}
\author{Jiong Cheng}
\affiliation{School of Physics and Optoelectronic Technology, Dalian University of
Technology, Dalian 116024,People's Republic of China}
\author{Wen-Dong Li}
\affiliation{Department of Physics, Ocean University of China, Qingdao 266100, People's
Republic of China}
\author{Ling Zhou}
\thanks{zhlhxn@dlut.edu.cn}
\affiliation{School of Physics and Optoelectronic Technology, Dalian University of
Technology, Dalian 116024,People's Republic of China}

\begin{abstract}
We propose a scheme in which the cooling of a mechanical resonator is
achieved by exposing the optomechanical system to a non-Markovian
environment. Because of the backflow from the non-Markovian environment, the
phonon number can go beyond the conventional cooling limit in a Markovian
environment. Utilizing the spectrum density obtained in the recent
experiment [Nature Communications \textbf{6}, 7606 (2015)], we show that the
cooling process is highly effective in a non-Markovian environment. The
analysis of the cooling mechanism in a non-Markovian environment reveals
that the non-Markovian memory effect is instrumental to the cooling process.
\end{abstract}

\maketitle


\address{\small $^1$School of Physics and Optoelectronic Technology, Dalian University of Technology, Dalian 116024,People's Republic of China\\
\small $^2$Department of Physics, Ocean University of China, Qingdao 266100, People's Republic of China}
\section{Introduction}
Recently, it has been widely recognized that optomechanical devices can be
used in detecting gravity waves \cite{PhysRevLett.113.151102,Abbott2009},
studying quantum-to-classical transitions \cite{PhysRevLett.112.080503},
performing high precision measurements \cite%
{PhysRevLett.114.113601,PhysRevLett.114.080503}, and processing quantum
information \cite{PhysRevA.91.063836,JPB.48.035503}. Cooling a mechanical
oscillator to its quantum ground state is a prerequisite for observing
quantum processes \cite{PhysRevLett.114.190502}. Several different kinds of
systems, for example, nanomechanical systems \cite%
{Jockel2014,PhysRevA.85.021802,Bagci2014}, micromechanical systems \cite%
{Nat.Commun.6.8491,Teufel2011}, and suspended mirrors in Fabry-P\'{e}rot
cavities \cite{Eerkens2015,Abbott2009} have been used for this purpose. For
all mechanical systems, thermal noise is unavoidable unlike other noise
sources that can be eliminated by using filters, screens, insulators,
\textit{etc}. It has been pointed out that the lowest phonon occupation
number $n_{f}$ in a mechanical oscillator is limited by the phonon number $%
n_{\text{th}}$ of the thermal environment \cite{RevModPhys.86.1391}. In
order to optimize the mechanical cooling, many methods have been proposed to
overcome the negative effects of the thermal environment, such as
dissipative cooling \cite{PhysRevLett.110.153606}, cooling through heat
pumping \cite{PhysRevLett.115.223602}, and cooling with mechanical
modulations \cite{PhysRevA.91.023818}. Up to today, studies on the
mechanical-oscillator cooling have all shown that the bath heating effect of
a mechanical oscillator can not be reversed easily in a typical dissipative
environment.\\
\hspace{-5mm} Most recently, a non-Markovian environment for a mechanical oscillator~\cite%
{Groblacher2015} was designed and its spectrum density was measured. A
non-Markovian environment exhibits the memory effect~\cite%
{PhysRevA.91.022328,PhysRevLett.105.240403,Reich2015} which may play a
positive role in cooling a mechanical oscillator. Therefore, it is of great
importance to develop a theoretical method to solve the problems related to
a nonlinear system in a non-Markovian environment.\\
\indent In this paper, we introduce a non-Markovian environment for a mechanical
oscillator. Taking the memory effect of the environment into consideration,
we obtain an analytical result for the dynamics of the phonon occupancy. We
then study the optimal cooling of an optomechanical system with different
spectrum densities $\mathcal{J}(\omega )$ through comparison with an
optomechanical system in a Markovian environment. Furthermore, this optimal
cooling can be realized with an experimental non-Ohmic spectrum density $%
\mathcal{J}(\omega )=C\omega ^{k}$, where $C>0$ and $k\in \mathbb{R}$~\cite%
{Groblacher2015}. To understand the mechanism of non-Markovian dynamics, we
analyze the equivalent energy-transport rate of the system in the cooling
process. The high-temperature environment can be regarded as a freezer for
the cooling of an oscillator when the non-Markovian memory effect is
included.
\section{Model and Hamiltonian}
\begin{figure}[tph]
\centering\includegraphics[width=8.5cm]{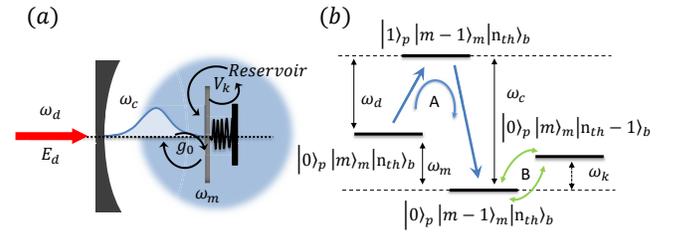} \vspace{0.2cm}
\caption{(Color online) (a) Typical optomechanical system coupled to a
general non-Markovian reservoir. (b) Schematic of the energy-level diagram
of the cavity-optomechanical system and its environment, where $|n\rangle
_{p}$, $|m\rangle _{m}$, and $|n_{\text{th}}\rangle _{b}$ represent the
number states of the cavity photons, the mechanical phonons, and the
reservoir phonons, respectively.}
\label{system}
\end{figure}

We consider a normal optomechanical system consisting of a cavity of
frequency $\omega _{c}$ and a mechanical resonator of frequency $\omega _{m}$
as shown in Fig.~\ref{system}(a).
The mechanical oscillator is coupled to a general non-Markovian reservoir~%
\cite{PhysRevA.83.032102,PhysRevA.81.052105} that can be realized with a
typical Nb point-contact superconducting quantum-interference device in an
LC circuit \cite{PhysRevB.30.1208} or a high-reflectivity Bragg mirror fixed
in the center of a doubly clamped Si$_{3}$N$_{4}$ beam in vacuum \cite%
{Groblacher2015}. The Hamiltonian of the system can be written as $%
H=H_{S}+H_{E}+H_{I}$, where
\begin{subequations}
\begin{eqnarray}
H_{S} &=&\hbar \omega _{c}a^{\dag }a+\hbar \omega _{m}b^{\dag }b-\hbar
g_{0}a^{\dag }a(b^{\dag }+b)  \notag \\
&&+i\hbar E(a^{\dag }e^{-i\omega _{d}t}-ae^{i\omega _{d}t}),
\label{eq:laser-coupling} \\
H_{E} &=&\sum_{k}\hbar \omega _{k}b_{k}^{\dag }b_{k}, \\
H_{I} &=&\sum_{k}\hbar V_{k}(b+b^{\dag })(b_{k}^{\dag }+b_{k}).
\label{eqsysi}
\end{eqnarray}%
Here $H_{S}$ describes the cavity mode driven by a laser coupled to the
mechanical resonator via radiation pressure with the coupling coefficient $%
g_{0}$ given by $g_{0}=(\omega _{c}/L)\sqrt{h/2m\omega _{m}}$. In Eq.~(\ref%
{eq:laser-coupling}), $\omega _{d}$ is the angular frequency of the laser
and $E$ is the cavity driving strength given by $E\equiv 2\sqrt{P\kappa
_{ex}/\hbar \omega _{d}}$ with $P$ the input power of the laser and $\kappa
_{ex}$ the input rate of the cavity. $H_{E}$ is the energy of the mechanical
reservoir with $\omega _{k}$ the frequency of the $k$th mechanical
oscillator. $H_{I}$ describes the coupling between the mechanical oscillator
and the reservoir with $V_{k}$ the coupling constant for the $k$th
environmental mode~\cite{PhysRevA.81.052105,PhysRevA.63.023812}.\\
\indent For the convenience of studying the effects in the cooling process, we transform the Hamiltonian into the displaced oscillator
representation in which the steady state of a cavity mode is the vacuum
state. As illustrated in Fig.~\ref{system}(b), the energy-level diagram is
constructed under the sideband-cooling condition $\omega _{c}=\omega
_{d}+\omega _{m}$. Kets $|0\rangle _{p}$, $|m\rangle _{m}$, and $|n_{\text{th%
}}\rangle _{b}$ are used to dentote respectively the number states of the
cavity, the mechanical oscillator, and the bath. Thus, we have anti-Stokes
processes \cite{PhysRevLett.99.093901} in which the transition $|1\rangle
_{p}|m\rangle _{m}|n_{\text{th}}\rangle _{b}\rightarrow |0\rangle
_{p}|m-1\rangle _{m}|n_{th}\rangle _{b}$ followed by the decay of the cavity
photon leads to cooling. In addition, under the rotating wave approximation
(RWA) with $\omega _{k}\approx \omega _{m}$, we can rewrite the Hamiltonian
for the system-reservoir coupling in Eq.~(\ref{eqsysi}) as $%
H_{I}=\sum_{k}\hbar V_{k}(bb_{k}^{\dag }+b^{\dag }b_{k})$ which allow the
mutual energy exchange processes $|0\rangle _{p}|m\rangle _{m}|n_{\text{th}%
}-1\rangle _{b}\leftrightarrow |0\rangle _{p}|m-1\rangle _{m}|n_{\text{th}%
}\rangle _{b}$ to occur in which the bidirectional action of the environment
can cool down or heat up the mechanical oscillator. Here the
anti-Stokes-like cooling process is not obvious. We will give a detailed
analysis of the cooling process in Section V to demonstrate clearly the
cooling mechanism of a non-Markovian environment.\\
\indent In the following calculations, we will go beyond the RWA and explore the
dynamics based on the full interaction Hamiltonian. With the full
Hamiltonian $H$ given in Eq.~(1), the Heisenberg-Langevin equations of
motion for the annihilation operators of the system are given by
\end{subequations}
\begin{subequations}
\begin{eqnarray}
\dot{a} &=&-(i\Delta _{c}+\frac{\kappa }{2})a+ig_{0}a(b+b^{\dag })+E+\sqrt{%
\kappa }a_{\text{in}}, \\
\dot{b} &=&-i\omega _{m}b+ig_{0}a^{\dag }a-i\sum_{k}V_{k}(b_{k}+b_{k}^{\dag
}),  \label{leqbd} \\
\dot{b_{k}} &=&-i\omega _{k}b_{k}-iV_{k}(b+b^{\dag }),  \label{leqbk}
\end{eqnarray}%
where $\Delta _{c}=\omega _{c}-\omega _{d}$ and $a_{\text{in}}$ is the input
noise operator of the cavity. For convenience, we take $\hbar =1$ in the
remaining part of the paper. The autocorrelation function of the vacuum
noise is $\langle a_{\text{in}}(t)a_{\text{in}}^{\dag }(\tau )\rangle
=\delta (t-\tau )$ \cite{QN}. Solving Eq.~(\ref{leqbk}) for the bath
operator $b_{k}(t)$, we have
\end{subequations}
\begin{eqnarray}
b_{k}(t) &=&b_{k}(0)e^{-i\omega _{k}t} \\
&&-iV_{k}\int_{0}^{t}d\tau \lbrack b(\tau )+b^{\dag }(\tau )]e^{-i\omega
_{k}(t-\tau )}.  \notag
\end{eqnarray}%
Substituting the above equation into Eq.~(\ref{leqbd}), we obtain
\begin{eqnarray}
\dot{b} &=&-i\omega _{m}b+ig_{0}a^{\dag }a  \label{eqb} \\
&&+\int_{0}^{t}d\tau f(t-\tau )[b(\tau )+b^{\dag }(\tau )]-\xi (t),  \notag
\end{eqnarray}%
where $f(t)=2i\sum_{k}V_{k}^{2}\sin (\omega _{k}t)=2i\int_{0}^{\infty
}d\omega \mathcal{J}(\omega )\sin (\omega t)$ with $\mathcal{J}(\omega )$
the spectral density of the reservoir. Instead of $[\xi (t),\xi (t^{\prime
})]\varpropto \delta (t^{\prime }-t)$ for a Markovian environment, the noise
operator $\xi (t)=i\sum_{k}V_{k}[b_{k}(0)e^{-i\omega _{k}t}+b_{k}^{\dag
}(0)e^{i\omega _{k}t}]$ has a non-local time correlation function for a
non-Markovian environment. We adopt the commonly-used spectral density of
the form $\mathcal{J}(\omega )=\eta \omega (\omega /\omega
_{0})^{s-1}e^{-\omega /\omega _{0}}$~\cite{RevModPhys.59.1}, where $\eta $
is the strength of the system-bath coupling and $\omega _{0}$ is the cut-off
frequency. The exponent $s$ is a real number that determines the $\omega $
dependence of $\mathcal{J}(\omega )$ in the low-frequency region. The baths
with $0<s<1$, $s=1$, and $s>1$ are referred to as the \textquotedblleft
sub-Ohmic\textquotedblright , the \textquotedblleft Ohmic\textquotedblright
, and the \textquotedblleft super-Ohmic\textquotedblright\ baths,
respectively. Here the memory kernel $f(t)$ characterizes the non-Markovian
dynamics of the reservoir.\\
\indent To study the dynamics of our system under the strong driving condition, we
make use of the linear approximation by decomposing the operators into the
classical and quantum components~\cite{RevModPhys.86.1391}, \textit{i.e.}, $%
a\rightarrow \alpha+a$ and $b\rightarrow \beta +b$. The time evolution of
the annihilation operators of the system in the Heisenberg picture is then
governed by
\begin{subequations}
\label{leq}
\begin{eqnarray}
\dot{\alpha} &=&-(i\Delta _{c}+\frac{\kappa }{2})\alpha +ig_{0}\alpha (\beta
+\beta ^{\ast })+E,  \label{leqa} \\
\dot{\beta} &=&-i\omega _{m}\beta +ig_{0}|\alpha |^{2}  \notag \\
&&+\int_{0}^{t}d\tau f(t-\tau )[\beta (\tau )+\beta ^{\ast }(\tau )],
\label{leqb} \\
\dot{a} &=&-(i\Delta _{c}^{\prime }+\frac{\kappa }{2})a+iG(b+b^{\dag })+%
\sqrt{\kappa }a_{\text{in}},  \label{leqc} \\
\dot{b} &=&-i\omega _{m}b+i(Ga^{\dag }+G^{\ast }a)  \notag \\
&&+\int_{0}^{t}d\tau f(t-\tau )[b(\tau )+b^{\dag }(\tau )]-\xi (t),
\label{leqd}
\end{eqnarray}%
where $\Delta _{c}^{\prime }(t)=\Delta _{c}-g_{0}[\beta (t)+\beta ^{\ast
}(t)]$ is the detuning modified by the optomechanical coupling and $%
G(t)=\alpha (t)g_{0}$ describes the linear coupling strength. We see that
the time-dependent coefficients $\Delta _{c}^{\prime }(t)$ and $G(t)$ are
determined by $\alpha (t)$ and $\beta (t)$. When the displacements of the
optical and mechanical modes, $\alpha (t)$ and $\beta (t)$, are large
enough, the linear approximation are satisfied. In the following
discussions, the choice of the parameters will ensure the validity of the
linear approximation.

\section{Dynamics of a mechanical oscillator}

In the system under study, the bath of the cavity mode is a Markovian
environment to ensure that the energy of an oscillator is taken away through
the cavity without reflux~\cite{PhysRevLett.99.093901}. To examine the
dynamics of a mechanical oscillator, we can eliminate the cavity mode by
solving Eq.~(\ref{leqa}) as follows
\end{subequations}
\begin{eqnarray}
a(t) &=&a(0)e^{u(t)} \\
&&+\int_{0}^{t}d\tau e^{u(t-\tau )}\{iG(\tau )[b(\tau )+b^{\dag }(\tau )]+%
\sqrt{\kappa }a_{\text{in}}(\tau )\},  \notag
\end{eqnarray}%
where $u(t_{1}-t_{2})=-\int_{t_{2}}^{t_{1}}d\tau \lbrack i\Delta ^{\prime
}(\tau )+\kappa /2]$ describes the effect of the equivalent detuning
resulting from the radiation pressure. We then have
\begin{eqnarray}
\dot{b} &=&-i\omega _{m}b+\int_{0}^{t}d\tau F(t-\tau )[b(\tau )+b^{\dag
}(\tau )]  \notag  \label{trb} \\
&&+A_{0}(t)+A_{\text{in}}(t)-\xi (t),
\end{eqnarray}%
where
\begin{eqnarray}
F(t-\tau ) &=&f(t-\tau )-[G^{\ast }(t)G(\tau )e^{u(t-\tau )}-H.c.],  \notag
\\
A_{0}(t) &=&i[G^{\ast }(t)e^{u(t)}a(0)+H.c.], \\
A_{\text{in}}(t) &=&\int_{0}^{t}d\tau i[\sqrt{\kappa }G^{\ast
}(t)e^{u(t-\tau )}a_{\text{in}}(\tau )+H.c.].  \notag
\end{eqnarray}%
The memory kernel $F(t)$ contains the effect of radiation pressure and $%
A_{0}(t)$ and $A_{\text{in}}(t)$ describe respectively the impact of the
initial condition $a(0)$ and the input noise $a_{\text{in}}(t)$.

We now focus on the mechanical oscillator. In consideration of the linearity
of Eq.~(\ref{trb}), we can assume that the solution of the operator $b(t\geq
0)$ is of the form
\begin{equation}
b(t)=M(t)b(0)+L^{\ast }(t)b^{\dag }(0)+S(t)  \label{b}
\end{equation}%
with the initial conditions $M(0)=1$ and $L(0)=0$. The equations for the
time-dependent coefficients $L(t)$, $M(t)$, and $S(t)$ can be found by
substituting Eq.~(\ref{b}) into Eq.~(\ref{trb}) and then comparing the
coefficients. We have
\begin{eqnarray}
\dot{M(t)} &=&-i\omega _{m}M(t)+\int_{0}^{t}d\tau F(t-\tau )[M({\tau }%
)+L(\tau )],  \notag  \label{mleq} \\
\dot{L(t)} &=&i\omega _{m}L(t)+\int_{0}^{t}d\tau F^{\ast }(t-\tau )[M({\tau }%
)+L(\tau )],  \notag \\
\dot{S(t)} &=&-i\omega _{m}S(t)+\int_{0}^{t}d\tau F(t-\tau )[S(\tau
)+S^{\dag }(\tau )]  \notag \\
&&+A_{0}(t)+A_{\text{in}}(t)-\xi (t).
\end{eqnarray}%
If $M(t)$ and $L(t)$ are known \cite{PhysRevA.81.052105}, the operator $S(t)$
can be completely determined through
\begin{eqnarray}
S(t) &=&\int_{0}^{t}d\tau \lbrack M(t-\tau )+L^{\ast }(t-\tau )]  \notag \\
&&\times \lbrack A_{0}(\tau )+A_{\text{in}}(\tau )-\xi (\tau )].
\end{eqnarray}

As shown in Eqs.~\eqref{b} and~\eqref{mleq}, the solutions of the quantum
parts are related to their classical parts. That is, the classical nonlinear
dynamics can be manifested in the quantum properties of the system~\cite%
{PhysRevLett.112.110406} even through we have made use of the linear
approximation. Especially, due to the memory effects in the non-Markovian
regime, the classical parts can not reach steady states as they can in the
Markovian regime~\cite{PhysRevLett.110.153606}. We will derive the dynamic
solutions of classical parts by solving Eqs.~\eqref{leqa} and \eqref{leqb}.
Substituting the time evolution of $\alpha (t)$ and $\beta (t)$ into Eqs.~%
\eqref{b}, \eqref{mleq}, and (11), we can obtain the time evolution of the
mechanical resonator. Hence, we can thoroughly solve the problem of the
non-Markovian dynamics of the phonon number without making any other
approximations except the linear approximation (see Appendix A for details).

\section{Sideband cooling in the Non-Markovian regime}
\begin{figure}[b]
\centering \includegraphics[width=8.5cm]{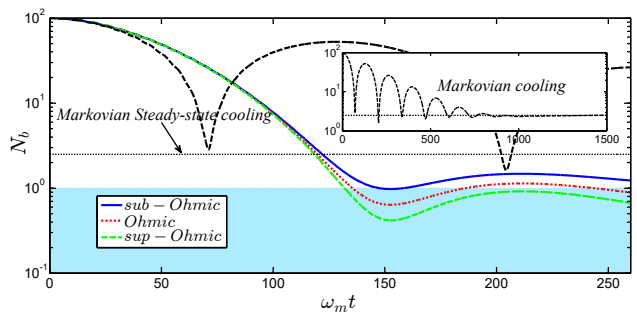}
\caption{(Color online) Time evolution of $N_{b}$ with a red detuning in
different regimes. The Markovian dissipation $\protect\gamma _{m}=10^{-8}%
\protect\omega _{m}$, $s=1/2$ for a sub-Ohmic bath, $s=1$ for a Ohmic bath, $%
s=2$ for a sup-Ohmic bath, the cutoff frequency $\protect\omega _{0}/\protect%
\omega _{m}=5$, and the coupling constant $\protect\eta =10^{-5}$ are used.
The values of other parameters are $\protect\kappa /\protect\omega _{m}=0.05$%
, $g_{0}/\protect\kappa =10^{-3}$, $E/\protect\omega _{m}=300$, $|\protect%
\alpha _{0}|=100$, $|\protect\beta_{0}|=100$, and $m_{0}=10^{2}$.}
\label{fig2}
\end{figure}

We now consider the non-Markovian effect for the sideband cooling with $%
\Delta ^{\prime }(t)=\omega _{m}$. In this case, we have $u(t)=-(i\omega
_{m}+\kappa /2)t$. To study the cooling dynamics in the non-Markovian
regime, we use Eq.~(\ref{b}) to obtain the time evolution of the mean phonon
number of the quantum part without initial system-reservoir correlations. We
assume that the initial quantum number distributions are given by $\langle
b^{\dag }(0)b(0)\rangle =m_{0}$, $\langle a^{\dag }(0)a(0)\rangle =n_{0}$, $%
\langle a_{in}(t)a_{in}^{\dag }(\tau )\rangle =\delta (t-\tau )$, and $%
\langle b_{k}^{\dag }(0)b_{k}(0)\rangle =m_{k}$ with $m_{k}=1/(e^{\hbar
\omega _{k}/k_{B}T}-1)$ the photon distribution function of the reservoir.
We set the mirror to be initially in thermal equilibrium with the
environment with $m_{0}=1/(e^{\hbar \omega _{m}/k_{B}T}-1)$. The time
evolution of the mean phonon number $N_{b}(t)$ is then given by
\begin{eqnarray}
N_{b}(t) &=&[|M(t)|^{2}+|L(t)|^{2}]m_{0}+|L(t)|^{2} \\
&&+\int_{0}^{t}\int_{0}^{t}d\tau _{1}d\tau _{2}[L(t-\tau _{1})+M^{\ast
}(t-\tau _{1})]  \notag \\
&&\times \lbrack L^{\ast }(t-\tau _{2})+M(t-\tau _{2})]  \notag \\
&&\times \lbrack f_{1}(\tau _{1},\tau _{2})+f_{2}(\tau _{1},\tau
_{2})+f_{3}(\tau _{1},\tau _{2})],  \notag
\end{eqnarray}%
where
\begin{eqnarray}
f_{1}(\tau _{1},\tau _{2}) &=&G(\tau _{1})G^{\ast }(\tau _{2})e^{-u(\tau
_{1}-\tau _{2})}n_{0} \\
&&+G^{\ast }(\tau _{1})G(\tau _{2})e^{u(\tau _{1}-\tau _{2})}(n_{0}+1),
\notag \\
f_{2}(\tau _{1},\tau _{2}) &=&|G(\tau _{1})|^{2}(1-e^{-\kappa \tau _{1}}),
\notag \\
f_{3}(\tau _{1},\tau _{2}) &=&\int_{0}^{\infty }\mathcal{J}(\omega )d\omega
\{e^{-i\omega (\tau _{1}-\tau _{2})}  \notag \\
&&+2\cos \omega (\tau _{1}-\tau _{2})(e^{\frac{\hbar \omega }{k_{B}T}%
}-1)^{-1}\}  \notag
\end{eqnarray}%
in which $f_{1}$ is the contribution from the cavity photons which depends
on the initial photon number $n_{0}$, $f_{2}$ results from the cavity input
noise, and $f_{3}$ represents the effect of the oscillator bath which
depends strongly on the spectral density $\mathcal{J}(\omega )$.

\begin{figure}[t]
\centering 
\includegraphics[width=8.5cm]{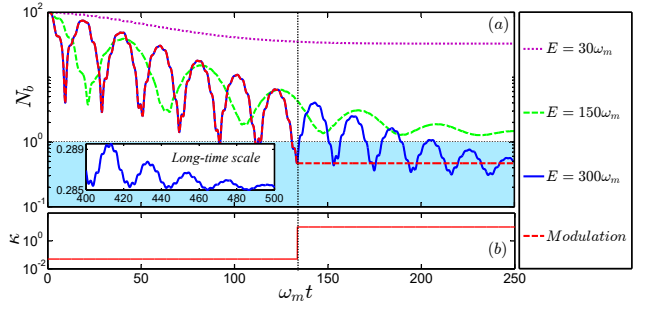}
\caption{(Color online) (a) Dynamics of the sideband cooling for different
values of the cavity driving strength $E$ with the cavity modulation ({\ the
red-dashed line}) $E/\protect\omega _{m}=300$. The inset shows the long-time
scale with the cavity driving strength given by $E/\protect\omega _{m}=300$.
The values of other parameters are the same as in Fig.~\protect\ref{fig2}
(b) Modulation scheme of the cavity dissipation rate $\protect\kappa $ for
fast cooling.}
\label{fig3}
\end{figure}

The time evolution of $N_{b}$ is depicted in Fig.~\ref{fig2} in which the
initial occupation number of the oscillator is chosen to be $m_{0}=100$.
For a typical suspended mirror optomechanical system with a frequency of the
order of $\omega _{m}=1\;\text{MHz}$, the bath temperature $T\approx 1.5\;%
\text{mK}$. In comparison, the bath temperature $T\approx 1.5\;\text{K}$ for
a typical optical micro-resonator with $\omega_{m}=1\;\text{GHz}$. We
compare the sideband cooling in the non-Markovian regime with several
different reservoir spectral densities $\mathcal{J}(\omega )$ with that in
the Markovian regime. For the Markovian regime, there exists a steady-state
cooling limit $n_{f}\approx \gamma_{m}n_{\text{th}}+n_{\text{ce}}$~\cite%
{RevModPhys.86.1391,PhysRevLett.99.093902}, where $n_{\text{th}}$ is the
equilibrium mechanical mode occupation number determined by the mechanical
bath temperature and $n_{\text{ce}}$ is positive and is determined by the
cavity mode and its environment.\\
\indent We first consider the Markovian regime. Because of the presence of the
Markovian reservoir, the mechanical oscillator is heated by its environment.
To highlight the contrastive results, we choose an extremely small value for
$\gamma _{m}$ with $\gamma _{m}=10^{-8}\omega _{m}$ in the Markovian regime
so that the negative effect of the Markovian environment of the oscillator
is negligible. See {the black-dashed line} in Fig.~\ref{fig2}.\\
\indent In the non-Markovian regime, the phonon number is quite different from that
in the Markovian regime. The time evolution of the phonon number for three
kinds of spectral densities is very similar in a short period of time with $%
\omega _{m}t<40$. However, as time develops, the memory effect gradually
sets in and the dynamics of the phonon number becomes different. Although
the asymptotic steady state can not be reached in the non-Markovian regime
due to the system-reservoir interaction, the mean phonon number, with small
vibrations, is far below that in the Markovian regime under extreme
conditions.\\
\indent Recently, the spectral density of a mechanical environment was measured
experimentally through the emitted light of a micro-optomechanical system~%
\cite{Groblacher2015}. The demonstration device consists of a thick layer of
Si$_{3}$N$_{4}$ with a high-reflectivity mirror pad at its center as a
mechanically moving end mirror in a Fabry-P\'{e}rot cavity. The spectral
density can be described by $\mathcal{J}(\omega )=C\omega ^{k}$ where $C>0$
and $k=-2.30\pm 1.05$. The concerned region of $\omega $, $\lbrack \omega _{%
\text{min}},\omega _{\text{max}}\rbrack$ with $\omega _{\text{min}}=885\;%
\text{kHz}$ and $\omega _{\text{max}}=945\;\text{kHz}$, is centered about
the mechanical resonance frequency $\omega _{m}=914\;\text{kHz}$ with the
bandwidth given by $\Gamma \approx 0.07\omega _{m}$. Utilizing the
experimental spectral density $\mathcal{J}(\omega )=C\omega ^{k}$ with $%
C=\eta e^{-\omega /\omega _{0}}/\omega _{0}^{k-1}$, $\Gamma _{m}$ $%
=0.1\omega_{m} $, and $k=-2$, we plot $N_{b}$ in Fig.~\ref{fig3} as a
function of time.\\
\indent From Fig.~\ref{fig3}, we see that the cavity driving strength $E$ affects
the cooling effect: The higher the value of $E$, the better the cooling
effect. The phonon number oscillates periodically for the narrow band
spectrum of the environment. After a long period of time, the mean phonon
number decreases and can reach the region with $N_{b}<0.3$ for $E=300\omega
_{m}$. In consideration of the practicability and feasibility, it is always
desired to speed up the cooling process. For this purpose, we can utilize
the Q-switch technology by increasing suddenly the loss rate when the mean
phonon number reaches the ideal value. As shown in Fig.~\ref{fig3} (see {the
solid-blue line}), the phonon occupation will reach a low-excitation level
in a non-steady state in a short period of time. For example, at $\omega
_{m}t=133.6$, $N_{b}\approx 0.46$. At this moment, we can accelerate the
stability of this low excitation state by increasing the damping rate $%
\kappa $ from $0.05\omega _{m}$ to $10\omega _{m}$. The modulation of the
cavity dissipation $\kappa $ can be realized by the Q-switch technology~\cite%
{PhysRevLett.110.153606}. The modulation results are shown in Fig.~\ref{fig3}%
(a) with {the red-dashed line}. The phonon number will attain a low and
stable value after modulation. The modulating signal is displayed in Fig.~%
\ref{fig3}(b). Furthermore, a comparison of {the dashed-black line} with {%
the solid-blue line} indicates that, for the same cavity driving strength $%
E=300\omega _{m}$, the cooling effect for a narrow-band spectrum~\cite%
{Groblacher2015} is smaller than that for the sup-Ohmic wide-band spectrum,
which implies that, with the current experimental technique for a
non-Markovian environment, we can obtain an optimal cooling effect.

\section{Non-Markovian cooling mechanism}

As we mentioned for Fig.~\ref{system}(b), a non-Markovian environment can
play two opposite roles: \textquotedblleft cooling\textquotedblright\ and
\textquotedblleft heating\textquotedblright . In order to understand the
cooling mechanism of a non-Markovian environment, we introduce the transport
rate $\upsilon _{j}=dN_{j}/dt$ with $j=a$ or $b$, where $N_{a}$ and $N_{b}$
are the occupation numbers of the cavity and the mechanical oscillator,
respectively. The differential equations for the mean values of the
second-order moments are given in Eqs.~(\ref{eq:dynamics-Na}) through~(\ref%
{eq:dynamics-bl_k_sq}) in Appendix B, in which only the beam splitter terms
survive under the condition that $g_{0}/\omega _{m},V_{k}/\omega
_{m},V_{k}/\omega _{k}\ll 1$. Then, we have $\upsilon _{a}=-\upsilon
_{\kappa }+\upsilon _{c}$ [see Eq.~(\ref{eq:dynamics-Na})], where $\upsilon
_{\kappa }=\kappa N_{a}$ represents the output flow of the photon energy
through the cavity dissipation and $\upsilon _{c}=i(G\langle a^{\dag
}b\rangle -G^{\ast }\langle ab^{\dag }\rangle )$ the input flow of the
photon energy resulting from the mechanical mode due to the optomechanical
interaction. The mechanical transport rate $\upsilon _{b}=-\upsilon
_{c}+\bigtriangleup \upsilon $ with $\bigtriangleup \upsilon
=i\sum_{k}V_{k}(\langle b^{\dag }b_{k}\rangle ^{\ast }-\langle b^{\dag
}b_{k}\rangle )$ [see Eq.~(\ref{eq:dynamics-Nb})] shows clearly that the
optomechanical coupling cools the mechanical oscillator. If we can achieve $%
\bigtriangleup \upsilon <0$, the mechanical oscillator will be cooled
further. Under the red-detuning condition with $\Delta ^{\prime }(t)\approx
\omega _{m}$, we have
\begin{equation}
\Delta \upsilon =2\text{Im}(\int_{0}^{t}d\tau \;i[F_{\text{th}%
}-\sum_{k}GV_{k}\langle a^{\dag }b_{k}\rangle e^{i\Delta _{k}(t-\tau )}]),
\label{eqdv}
\end{equation}%
where $\Delta _{k}=\omega _{m}-\omega _{k}$ and $F_{\text{th}%
}=\int_{0}^{\infty }d\omega \mathcal{J}(\omega )[N_{\omega
}-N_{b}]e^{i(\omega _{m}-\omega )(t-\tau )}$ with $N_{\omega }=(e^{\hbar
\omega /k_{B}T}-1)^{-1}$ describes the heat conduction effect from the
mechanical reservoir to the oscillator. For the cooling of the mechanical
oscillator, the phonon number $N_{b}$ of the mechanical oscillator is
usually smaller than the phonon number $N_{\omega }$ of its environment. The
integral\ $\int_{0}^{\infty }d\omega \mathcal{J}(\omega )e^{i(\omega
_{m}-\omega )(t-\tau )}$ represents the equivalent damping rate. If $%
\mathcal{J}(\omega )$ is a flat spectrum, then $\int_{0}^{\infty }d\omega
\mathcal{J}(\omega )e^{i(\omega _{m}-\omega )(t-\tau )}=\gamma _{m}/2$. We
can thus infer that the equivalent damping rate $\int_{0}^{\infty }d\omega
\mathcal{J}(\omega )e^{i(\omega _{m}-\omega )(t-\tau )}$ is always positive.
Hence, $F_{\text{th}}$ makes a positive contribution to $\Delta \upsilon $
because it is just the integral of $\mathcal{J}(\omega )e^{i(\omega
_{m}-\omega )(t-\tau )}$ with the positive weight $N_{\omega }-N_{b}$. We
can therefore conclude that the heat conduction of the bath has a negative
effect for the cooling because of the higher thermal occupation of the
environment in the cooling process.

If the second term $\sum_{k}GV_{k}\langle a^{\dag }b_{k}\rangle e^{i\Delta
_{k}(t-\tau )}$\ in Eq.~(14) has a positive value so as to compensate the
first term, we can have $\Delta \upsilon <0$ so that we can achieve cooling.
Of course, it can have a negative value or even a complex value. For a
negative value, the non-Markovian backflow contributes even a worse effect
for the cooling than that in a Markovian environment. In reference to {the
purple-dashed line} with $E=30\omega _{m}$ in Fig.~\ref{fig3}, the value of $%
N_{b}$ is even higher than that in the Markovian case for $E=300\omega _{m}$
\cite{note} (see {the purple-dashed line} in Fig.~\ref{fig2}). For a
Markovian reservoir with no backflow, $\sum_{k}GV_{k}\langle a^{\dag
}b_{k}\rangle e^{i\Delta _{k}(t-\tau )}$ is zero. Therefore, the total heat
conduction effect can be described as $\Delta \upsilon =$ $\gamma _{m}n_{%
\text{th}}>0$. In other words, a Markovian reservoir will definitely have a
negative effect on the cooling if it is desired that the temperature of the
mechanical oscillator is lower than that of its environment. According to
Eq.~(\ref{eqdv}), if the second term is greater than the first one, then we
can have a further net cooling effect. In other words, \textquotedblleft
cooling\textquotedblright\ the mechanical oscillator in the ideal case
depends on the second term. The larger the second term, the better the
cooling effect.\\
\indent From the form of the second term, we can draw three conclusions: (1) The
linearized coupling coefficient $G=\alpha g_{0}$ is a controllable
parameter. We can increase $\alpha $ by enhancing the cavity driving
strength to achieve the ideal cooling effect, which is exactly what is
demonstrated in Fig.~\ref{fig3}. Using the parameter $\alpha $ to speed up
the cooling process was discussed by Liu \emph{et al.} \cite%
{PhysRevLett.110.153606}, which is in consistency with Eq.(14). (2) The
factor $\langle a^{\dag }b_{k}\rangle $ indicates clearly that the backflow
from the non-Markovian environment into the cavity field via the mechanical
oscillator does indeed cool the mechanical oscillator with the processes $%
\langle b^{\dag }b_{k}\rangle $ and $\langle a^{\dag }b\rangle $ involved.
Certainly, the Markovian environment of the cavity field is still needed
since it is the final place for the lost energy of the mechanical
oscillator. (3) The frequency component $\omega _{k}=\omega _{m}$ $(\Delta
_{k}=0)$ yields the main contribution. If $\Delta _{k}\gg GV_{k}$, the
second term in Eq.~(14) is fast oscillating and makes no contribution to the
cooling. In order to maintain the optimal cooling effect, the frequency of
the environment should be centered about the frequency of the mechanical
oscillator. Hence, when the non-Markovian memory effect is included, even
the temperature of the bath is much higher than that of the phonon mode, and
the environment could be also regarded as a freezer in the cooling of the
oscillator.\\
\indent In the review process of the paper, we noticed the new publication~\cite%
{cooling} in which the ultrafast optimal sideband cooling with a
non-Markovian evolution is proposed. They optimally designed the coupling
functions so as to optimize the cooling rate in both Markovian and
non-Markovian environments for the cavity field as well as for the
mechanical oscillator. Different from the work reported in~\cite{cooling},
we aim to achieve lower phonon numbers in the long-time scale. Through the
analysis of the cooling mechanism in a non-Markovian environment, we showed
that the backflow from the non-Markovian environment of the mechanical
oscillator into the cavity field with the further decay into the non-memory
environment of the cavity field is the cause for lower phonon numbers. This
conclusion coincides with \cite{cooling} in which the non-Markovian dynamics
in the cavity field deteriorates their cooling protocol.

\section{conclusions}

In this paper, we put forward an environment engineering scheme for the
mechanical cooling. Making use of several spectra including the one obtained
in the experiment \cite{Groblacher2015}, we showed that the cooling effect
in the present scheme is better than that in a Markovian environment. We
also analyzed the cooling mechanism of a non-Markovian environment. Our
analysis showed that the mechanical oscillator environment is not always
detrimental to the cooling. If the environment possesses the non-Markovian
memory effect, not only is the entanglement \cite%
{PhysRevA.91.022328,Reich2015} protected but also the mechanical cooling is
optimized. A high temperature bath could be also regarded as a freezer to
the cooling of the oscillator. Most importantly, with the present experiment
technology, we can use a non-Markovian environment to cool a mechanical
oscillator so as to go beyond the limit of a Markovian environment.\\
\indent We would like to thank Mr. Wen-Lin Li and Tian-Yi Chen for helpful
discussions. We also appreciate Prof. Fuxiang Han for his polishing in
language. This work was supported by the NSF of China under Grant No.
11474044.
\begin{widetext}
\appendix
\section{DYNAMICS OF A MECHANICAL OSCILLATOR}
From the expressions of $A_{0}$, $A_{\text{in}}$, and $\xi $, we can find that $\langle A_{0}(t)A_{\text{in}}(\tau )\rangle =0$, $\langle A_{0}(t)\xi (\tau )\rangle =0$, and $\langle A_{\text{in}}(t)\xi (\tau )\rangle =0$. The phonon number $N_{b}$ of the quantum part reads
\begin{eqnarray}
\langle b^{\dag }(t)b(t)\rangle  &=&|M(t)|^{2}\langle b^{\dag
}(0)b(0)\rangle +|L(t)|^{2}\langle b(0)b^{\dag }(0)\rangle +M^{\ast
}(t)\langle b^{\dag }(0)S(t)\rangle +M(t)\langle S^{\dag }(t)b(0)\rangle
\notag  \label{nbeq} \\
&&+L(t)\langle b(0)S(t)\rangle +L^{\ast }(t)\langle S^{\dag }(t)b^{\dag
}(0)\rangle +\langle S^{\dag }(t)S(t)\rangle ,
\end{eqnarray}%
where
\begin{subequations}
\begin{eqnarray}
\langle b^{\dag }(0)S(t)\rangle  &=&\int_{0}^{t}d\tau \lbrack M(t-\tau
)-L^{\ast }(t-\tau )]\langle b^{\dag }(0)[A_{0}(\tau )-\xi (\tau )]\rangle ,
\\
\langle b(0)S(t)\rangle  &=&\int_{0}^{t}d\tau \lbrack M(t-\tau )-L^{\ast
}(t-\tau )]\langle b(0)[A_{0}(\tau )-\xi (\tau )]\rangle , \\
\langle S^{\dag }(t)b(0)\rangle  &=&\int_{0}^{t}d\tau \lbrack L(t-\tau
)-M^{\ast }(t-\tau )]\langle \lbrack A_{0}(\tau )-\xi (\tau )]b(0)\rangle ,
\\
\langle S^{\dag }(t)b^{\dag }(0)\rangle  &=&\int_{0}^{t}d\tau \lbrack
L(t-\tau )-M^{\ast }(t-\tau )]\langle \lbrack A_{0}(\tau )-\xi (\tau
)]b^{\dag }(0)\rangle , \\
\langle S^{\dag }(t)S(t)\rangle  &=&\int_{0}^{t}\int_{0}^{t}[L(t-\tau
_{1})-M^{\ast }(t-\tau _{1})][M(t-\tau _{2})-L^{\ast }(t-\tau _{2})]  \notag
\\
&&\times \lbrack \langle A_{0}(\tau _{1})A_{0}(\tau _{2})\rangle +\langle
A_{in}(\tau _{1})A_{in}(\tau _{2})\rangle +\langle \xi (\tau _{1})\xi (\tau
_{2})\rangle ]
\end{eqnarray}%
in which the autocorrelation functions are given by
\end{subequations}
\begin{eqnarray}
\langle A_{0}(\tau _{1})A_{0}(\tau _{2})\rangle  &=&-[G(\tau _{1})G^{\ast
}(\tau _{2})e^{u^{\ast }(\tau _{1})+u(\tau _{2})}\langle a^{\dag
}(0)a(0)\rangle +G^{\ast }(\tau _{1})G(\tau _{2})e^{u(\tau _{1})+u^{\ast
}(\tau _{2})}\langle a(0)a^{\dag }(0)\rangle ],  \notag \\
\langle A_{in}(\tau _{1})A_{in}(\tau _{2})\rangle  &=&-|G(\tau
_{1})|^{2}\kappa e^{u(\tau _{1})+u^{\ast }(\tau _{1})}\int_{0}^{\tau
_{1}}d\tau e^{-u(\tau )-u^{\ast }(\tau )}\langle a_{in}(\tau )a_{in}^{\dag
}(\tau )\rangle ,  \notag \\
\langle \xi (\tau _{1})\xi (\tau _{2})\rangle
&=&-\sum_{k}V_{k}^{2}[e^{-i\omega _{k}(\tau _{1}-\tau _{2})}\langle
b_{k}(0)b_{k}^{\dag }(0)\rangle +e^{i\omega _{k}(\tau _{1}-\tau
_{2})}\langle b_{k}^{\dag }(0)b_{k}(0)\rangle ].
\end{eqnarray}%
The cross-correlation functions are given by
\begin{eqnarray}
\langle b^{\dag }(0)A_{0}(t)\rangle  &=&i[G^{\ast }(t)e^{u(t)}\langle
b^{\dag }(0)a(0)\rangle +G(t)e^{u^{\ast }(t)}\langle a(0)b(0)\rangle ^{\ast
}],  \notag \\
\langle b^{\dag }(0)\xi (t)\rangle  &=&i\sum_{k}V_{k}[e^{-i\omega
_{k}t}\langle b_{k}(0)b^{\dag }(0)\rangle ^{\ast }+e^{i\omega _{k}t}\langle
b_{k}(0)b(0)\rangle ^{\ast }],  \notag \\
\langle b(0)A_{0}(t)\rangle  &=&i[G^{\ast }(t)e^{u(t)}\langle
b(0)a(0)\rangle +G(t)e^{u^{\ast }(t)}\langle a(0)b^{\dag }(0)\rangle ^{\ast
}],  \notag
\end{eqnarray}
\begin{eqnarray}
\langle b(0)\xi (t)\rangle  &=&i\sum_{k}V_{k}[e^{-i\omega _{k}t}\langle
b(0)b_{k}(0)\rangle +e^{i\omega _{k}t}\langle b(0)b_{k}^{\dag }(0)\rangle ],
\notag \\
\langle A_{0}(t)b(0)\rangle  &=&i[G^{\ast }(t)e^{u(t)}\langle
a(0)b(0)\rangle +G(t)e^{u^{\ast }(t)}\langle b^{\dag }(0)a(0)\rangle ^{\ast
}],  \notag \\
\langle \xi (t)b(0)\rangle  &=&i\sum_{k}V_{k}[e^{-i\omega _{k}t}\langle
b_{k}(0)b(0)\rangle +e^{i\omega _{k}t}\langle b_{k}^{\dag }(0)b(0)\rangle ],
\notag \\
\langle A_{0}(t)b^{\dag }(0)\rangle  &=&i[G^{\ast }(t)e^{u(t)}\langle
a(0)b^{\dag }(0)\rangle +G(t)e^{u^{\ast }(t)}\langle b(0)a(0)\rangle ^{\ast
}],  \notag \\
\langle \xi (t)b^{\dag }(0)\rangle  &=&i\sum_{k}V_{k}[e^{-i\omega
_{k}t}\langle b(0)b_{k}^{\dag }(0)\rangle ^{\ast }+e^{i\omega _{k}t}\langle
b(0)b_{k}(0)\rangle ^{\ast }].
\end{eqnarray}%
The solution to Eq.~(\ref{nbeq}) depends on the initial values of the photon-phonon
correlation functions $\langle a(0)b(0)\rangle $, $\langle b(0)a(0)\rangle $, $\langle
a(0)b^{\dag }(0)\rangle $, and $\langle b^{\dag }(0)a(0)\rangle $ and on the initial values of the
mirror-reservoir correlation functions $\langle b(0)b_{k}(0)\rangle $, $\langle
b_{k}(0)b(0)\rangle $, $\langle b(0)b_{k}^{\dag }(0)\rangle $, and $\langle
b_{k}^{\dag }(0)b(0)\rangle $. Equation~\eqref{nbeq} can be solved by making use of
Eq.~\eqref{mleq} and the initial values of the system-reservoir correlation functions.

\section{DYNAMICS OF THE SYSTEM}

To understand clearly the role played by the environment in the cooling process, we study the time evolution of
the photon and phonon numbers by making use of the original Heisenberg-Langevin equations. Applying the linear approximation, we can simplify the dynamical equations.
Here, the system is surrounded by a large environment whose occupation
number $N_{k}$ can be regarded as a constant. The simplified dynamical equations are given by
\begin{subequations}
\label{single}
\begin{eqnarray}
\frac{d\langle N_{a}\rangle }{dt} &=&-\kappa N_{a}+i(G\langle a^{\dag
}b\rangle -G^{\ast }\langle a^{\dag }b\rangle ^{\ast }+G\langle ab\rangle
^{\ast }-G^{\ast }\langle ab\rangle ),\label{eq:dynamics-Na}\\
\frac{d\langle N_{b}\rangle }{dt} &=&-i(G\langle a^{\dag }b\rangle -G^{\ast
}\langle a^{\dag }b\rangle ^{\ast }+G\langle ab\rangle ^{\ast }-G^{\ast
}\langle ab\rangle )+i\sum_{k}V_{k}(\langle bb_{k}\rangle -\langle
bb_{k}\rangle ^{\ast }+\langle b^{\dag }b_{k}\rangle ^{\ast }-\langle
b^{\dag }b_{k}\rangle ),\label{eq:dynamics-Nb}\\
\frac{d\langle a^{\dag }b\rangle }{dt} &=&-[i(\omega _{m}-\Delta
_{c}^{\prime })+\kappa /2]\langle a^{\dag }b\rangle -i(G^{\ast }\langle
b^{2}\rangle +G^{\ast }N_{b}-G^{\ast }N_{a}-G\langle a^{2}\rangle ^{\ast
})-i\sum_{k}V_{k}(\langle a^{\dag }b_{k}\rangle +\langle ab_{k}\rangle
^{\ast }), \\
\frac{d\langle a^{\dag }b_{k}\rangle }{dt} &=&-[i(\omega _{k}-\Delta
_{c}^{\prime })+\kappa /2]\langle a^{\dag }b_{k}\rangle -iG^{\ast }(\langle
bb_{k}\rangle +\langle b^{\dag }b_{k}\rangle )-iV_{k}(\langle a^{\dag
}b\rangle +\langle ab\rangle ^{\ast }), \\
\frac{d\langle b^{\dag }b_{k}\rangle }{dt} &=&i(\omega _{m}-\omega
_{k})\langle b^{\dag }b_{k}\rangle -i(G^{\ast }\langle ab_{k}\rangle
+G\langle a^{\dag }b_{k}\rangle )+iV_{k}(\langle b_{k}^{2}\rangle
+N_{k}-N_{b}-\langle b^{2}\rangle ^{\ast }), \\
\frac{d\langle ab\rangle }{dt} &=&-[i(\Delta _{c}^{\prime }+\omega
_{m})+\kappa /2]\langle ab\rangle +i(G\langle b^{2}\rangle +GN_{b}+G^{\ast
}\langle a^{2}\rangle +G\langle aa^{\dag }\rangle )-i\sum_{k}V_{k}(\langle
ab_{k}\rangle +\langle a^{\dag }b_{k}\rangle ^{\ast }), \\
\frac{d\langle ab_{k}\rangle }{dt} &=&-[i(\Delta _{c}^{\prime }+\omega
_{k})+\kappa /2]\langle ab_{k}\rangle +iG(\langle bb_{k}\rangle +\langle
b^{\dag }b_{k}\rangle )-iV_{k}(\langle ab\rangle +\langle a^{\dag }b\rangle
^{\ast }), \\
\frac{d\langle bb_{k}\rangle }{dt} &=&-i(\omega _{m}+\omega _{k})\langle
bb_{k}\rangle +i(G^{\ast }\langle ab_{k}\rangle +G\langle a^{\dag
}b_{k}\rangle )-iV_{k}(\langle b_{k}^{2}\rangle +N_{k}+\langle b^{2}\rangle
+\langle bb^{\dag }\rangle ), \\
\frac{d\langle a^{2}\rangle }{dt} &=&-(2i\Delta _{c}^{\prime }+\kappa
)\langle a^{2}\rangle +2iG(\langle ab\rangle +\langle a^{\dag }b\rangle
^{\ast }), \\
\frac{d\langle b^{2}\rangle }{dt} &=&-2i\omega _{m}\langle b^{2}\rangle
+2i(G^{\ast }\langle ab\rangle +G\langle a^{\dag }b\rangle
)-2i\sum_{k}V_{k}(\langle bb_{k}\rangle +\langle b^{\dag }b_{k}\rangle
^{\ast }), \\
\frac{d\langle b_{k}^{2}\rangle }{dt} &=&-2i\omega _{k}\langle
b_{k}^{2}\rangle -2iV_{k}(\langle bb_{k}\rangle +\langle b^{\dag
}b_{k}\rangle ).\label{eq:dynamics-bl_k_sq}
\end{eqnarray}%
\end{subequations}
The equivalent transport rates of the cavity and the mechanical modes are $\upsilon
_{a}=dN_{a}/dt$ and $\upsilon _{b}=dN_{b}/dt$, respectively.
We set the laser to be red detuned with $\Delta ^{\prime }(t)=\omega _{m}$ for which
the beam splitter interaction is in resonance. Under the weak-coupling
condition that $g_{0}/\omega _{m}, V_{k}/\omega _{m}, V_{k}/\omega _{k}\ll 1$, we can identify the fast oscillating terms with the corresponding evolution equations given in Eqs.~(\ref{single}f) through~(\ref{single}%
k). We can then adiabatically eliminate these fast oscillating terms in Eqs.~(\ref{single}a) through~(\ref{single}e). In Eq.~(\ref{single}a), the first term contains the transport rate through the dissipation rate $\kappa $ and the second term contains the transport rate $\upsilon _{c}$ with $\upsilon _{c}=i(G\langle a^{\dag
}b\rangle -G^{\ast }\langle ab^{\dag }\rangle )$ from the mechanical mode through the optomechanical coupling. In Eq.~(\ref{single}b), the first term contains the output flow $\upsilon _{c}$ through the
optomechanical coupling and the second term contains the transport rate $\bigtriangleup \upsilon =i\sum_{k}V_{k}(\langle b^{\dag }b_{k}\rangle ^{\ast }-\langle b^{\dag }b_{k}\rangle )$ due to the oscillator-reservoir interaction. Combining Eqs.~(\ref{single}b) and~(\ref{single}e) we have
\begin{equation}
\bigtriangleup \upsilon =2V_{k}\text{Im}(\int_{0}^{t}d\tau\;
i[\sum_{k}e^{i\Delta _{k}(t-\tau )}V_{k}^{2}(N_{k}-N_{b})-\sum_{k}e^{i\Delta
_{k}(t-\tau )}GV_{k}\langle a^{\dag }b_{k}\rangle ]),  \label{eq:bdbk}
\end{equation}%
where $\Delta _{k}=\omega _{m}-\omega _{k}$ and $\sum_{k}e^{i\Delta _{k}(t-\tau
)}V_{k}^{2}(N_{k}-N_{b})=\int_{0}^{\infty }d\omega \mathcal{J}(\omega )[(e^{\hbar \omega/k_{B}T}-1)^{-1}-N_{b}]e^{i(\omega _{m}-\omega )(t-\tau
)}$ describes the heat conduction effect from the mechanical reservoir to the oscillator. The second term in the integrand in \eref{eq:bdbk} that contains $\langle a^{\dag }b_{k}\rangle $ describes the non-Markovian memory effect from the cavity dynamics through the optomechanical interaction. The two terms in the integrand in \eref{eq:bdbk} make opposite contributions in
the cooling process. When $\bigtriangleup \upsilon <0$, the reservoir will
exhibit a \textquotedblleft cooling\textquotedblright\ effect to the mechanical
oscillator. We can achieve this effect by increasing the linearized coupling rate
$G$. We also notice that, if $\Delta _{k}\gg GV_{k}$, the second term then becomes
fast oscillating and hence makes no contribution to the cooling. To maintain the
cooling effect, we should have $\Delta _{k}\ll GV_{k}$. That is, the frequency of the environment should be centered about the frequency of the mechanical oscillator. Therefore, when the system exhibits non-equilibrium dynamics~\cite{arXiv.1511.01267}, it is possible that the temperature of the mechanical oscillator is much
lower than that of the environment.
\newline
\end{widetext}%



\begin{references}
\bibitem{note} In Markovian regime, the the quantum cooling limit $N_{limit}\propto|G|^2$. The cooling limit for smaller $E=30\omega_m$ is lower  than that for $E=300\omega_m$.
\end{references}
\end{document}